# Toward a System Building Agenda for Data Integration

AnHai Doan, Adel Ardalan, Jeffrey R. Ballard, Sanjib Das
Yash Govind, Pradap Konda, Han Li, Erik Paulson, Paul Suganthan G.C., Haojun Zhang
University of Wisconsin-Madison

## ABSTRACT

In this paper we argue that the data management community should devote far more effort to building data integration (DI) systems, in order to truly advance the field. Toward this goal, we make three contributions. First, we draw on our recent industrial experience to discuss the limitations of current DI systems. Second, we propose an agenda to build a new kind of DI systems to address these limitations. These systems guide users through the DI workflow, step by step. They provide tools to address the "pain points" of the steps, and tools are built on top of the Python data science and Big Data ecosystem (PyData). We discuss how to foster an ecosystem of such tools within PyData, then use it to build DI systems for collaborative/cloud/crowd/lay user settings. Finally, we discuss ongoing work at Wisconsin, which suggests that these DI systems are highly promising and building them raises many interesting research challenges.

## 1. INTRODUCTION

Data integration (DI) has been an important research area in data management, and will become even more so in the age of Big Data and data science. Most DI works so far have focused on developing algorithms [4].

Going forward we argue that far more effort should be devoted to building DI systems. DI is engineering by nature. We cannot just keep developing DI algorithms in a vacuum. At some point we need to build systems to evaluate these algorithms, to integrate disparate R&D efforts, and to make practical impacts.

In this aspect, DI can take inspiration from RDBMSs and Big Data systems. Pioneering systems such as System R, Ingres, and Hadoop have really helped push these fields forward, by helping to evaluate research ideas, providing an architectural blueprint for the entire community to focus on, facilitating more advanced systems, and making widespread real-world impacts.

The question then is what kinds of DI systems we should build, and how? In this paper we make three contributions toward answering this question. First, in the past few years we have been working extensively in industry, using off-the-shelf DI systems as well as building new systems to address DI problems in social media, Web, e-commerce, and the Internet of Buildings. Based on this experience, we discuss the limitations of current DI systems that we believe prevent them from being used widely in practice. Among others, these limitations include not providing support for the entire DI pipeline, not providing detailed guidance for human users, not addressing the true pain points of the DI process, and being built as monolithic stand-alone systems making it very difficult to extend and exploit necessary techniques (e.g., visualization, learning, crowdsourcing, etc.).

Second, we propose a novel agenda to build a new kind of DI systems to address the above limitations. In contrast to current systems that often seek to *automate* the entire DI pipeline, these new systems assume the *human user* drives the DI process. The new systems provide detailed *how-to guides* to help the user through this process, step by step, and provide automated tools to address the "pain points" of the steps. These guides and tools seek to cover the *entire* DI process, not just a few steps as current DI systems typically do. Finally, tools are built on top of the Python data science and Big Data ecosystem (PyData), and thus can easily exploit a wide range of techniques, e.g., visualization, learning, extraction, cleaning, SQL querying, etc. We discuss how to foster PyDI, an ecosystem of such open-source DI tools, as a part of PyData, then use it to build DI systems for collaborative/cloud/crowd/lay user settings. As an example of such systems, we discuss "hands-off" DI systems, which solve the entire DI task using only crowdsourcing.

Finally, we describe initial work on this agenda at Wisconsin and lessons learned. That work currently focuses on building DI systems for entity matching (EM) [10], string similarity joins, attribute value normalization/verification, and on building hands-off DI systems on the cloud [7]. Several systems have been released (and used extensively by industrial partners). Our experience so far suggests that these systems are highly promising and building them raises numerous interesting research challenges.

**Related Work:** As far as we can tell, no broad DI system building agendas have been proposed, though interesting future directions were discussed in [13, 1, 4, 6]. Recent surveys and textbooks include [4, 5, 9, 1], and recent pioneering projects are discussed in [13, 2, 8, 6, 11].

Perhaps the work closest to ours is [12], which proposes OpenII, an open-source DI platform. That work however does not discuss a system building agenda, and the platform is not built on a data science stack, as ours does. Recently we have also applied the agenda proposed here to the context of EM, building Magellan, a general-purpose EM system [10] and Corleone and Falcon, hands-off crowdsourced EM systems [7, 3]. Those works focus only on EM, whereas this paper discusses issues general to the entire DI field. They also do not discuss fostering an ecosystem of DI tools, nor consider collaborative/cloud/crowd/lay user settings, as this paper does.

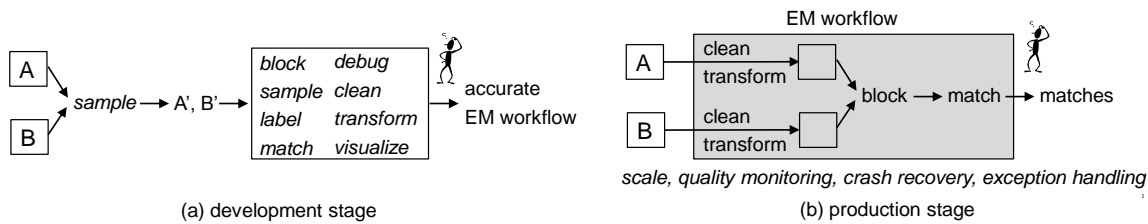

Figure 1: Matching two tables in practice often involves two stages and many steps (shown in italics).

## 2. MOTIVATING EXAMPLE

To make subsequent discussions concrete, in this section we will describe a running example. This example focuses on entity matching (EM), and shows that users often do EM in two stages, development and production, using many steps. This example is selected because EM has been a major focus of the ongoing work at Wisconsin. But it is also quite representative of many other DI tasks (e.g., wrapper-based extraction, schema matching, data cleaning).

EXAMPLE 1. *Consider matching two tables A and B each having 1M tuples, i.e., find all pairs ($a \in A, b \in B$) that refer to the same real-world entity. In practice, a user U typically solves this task in two stages: development and production.*

*In the development stage (Figure 1.a), U tries to find an accurate EM workflow. This is often done using data samples (because working directly with the large tables A and B is very time consuming, especially given the iterative nature of this stage). Specifically, U first samples two smaller tables $A'$ and $B'$ (each having 100K tuples, say) from A and B. Next, U performs blocking on $A'$ and $B'$ to remove obviously non-matched tuple pairs. U often must try and debug different blocking techniques to find the best one.*

*Suppose U wants to apply supervised learning to match the tuple pairs that survive the blocking step. Then next, U may take a sample S from the set of such pairs, label pairs in S (as matched / non-matched), and then use the labeled sample to develop a learning-based matcher (e.g., a classifier). U often must try and debug different learning techniques to develop the best matcher.*

*Once U is satisfied with the accuracy of the matcher, the production stage begins (Figure 1.b). In this stage, U executes the EM workflow that consists of the developed blocking strategy followed by the matcher on the original tables A and B. To scale, U may need to rewrite the code for blocking and matching to use Hadoop or Spark.*

## 3. LIMITATIONS OF CURRENT SYSTEMS

Current DI systems fall into two main groups, depending on whether they try to solve just *a single DI problem* (e.g., EM, schema matching), or to jointly solve *multiple DI problems* (e.g., schema matching, followed by schema integration, then EM). We now discuss these two groups in turn.

### 3.1 Systems for a Single DI Problem

Our experience suggests that systems in this group suffer from the following limitations that prevent them from being used extensively in practice.

**1. Do Not Solve All Steps of the DI Task:** When solving a DI task users often must execute many steps, e.g., sampling, blocking, labeling, matching, debugging, etc. (see Example 1). Current systems provide support for only a few steps (e.g., blocking, matching), ignoring less well-known yet equally critical steps (e.g., sampling, labeling, debugging). One may think that the ignored steps are just trivial engineering. Yet in practice this is anything but.

EXAMPLE 2. *Consider the sampling and labeling steps in Example 1. Let C be the set of tuple pairs surviving blocking. If C contains relatively few matches (a common situation in practice), then a random sample S from C may contain few if any matches, thus is not suitable for training a matcher. In such cases how should user U take a sample S from C?*

*After sampling, labeling tuple pairs in sample S as matched or non-matched seems trivial. Yet it is actually quite complicated in practice. Very often, during the labeling process user U gradually realizes that his/her current definition of what it means to be a match is incorrect or inadequate. Revising this definition however requires U to revisit and potentially relabel pairs that have already been labeled, a very tedious and time-consuming process.*

As yet another example, how to debug the blocking/matching steps has proven to be quite difficult in practice.

**2. Difficult to Exploit a Wide Range of Techniques:** Each of the above steps often exploit many techniques, e.g., SQL querying, keyword search, learning, visualization, information extraction (IE), outlier detection, crowdsourcing, etc. Today, however, it is very difficult to exploit a wide range of such techniques. Incorporating all such techniques into a single DI system has proven highly challenging.

The alternative solution of moving data among multiple systems, e.g., an EM system, an IE system, a visualization system, etc., also does not work. This is because solving a DI task is often an iterative process. So we would end up moving data among multiple systems *repeatedly*, often by reading/writing to disk and translating among proprietary data formats numerous times, in a tedious and time consuming process. A fundamental problem here is that most current DI systems are *stand-alone monoliths* that are not designed from scratch to "play well" with other systems.

**3. Little Guidance for Users on Solving the DI Task:** In many DI scenarios users often do not know what steps to take, in what order. For example, suppose a user U wants to perform EM with at least 95% precision and 80% recall. How should U start? Should U use a learning-based or a rule-based EM approach? What should U do if after many tries U still cannot reach 80% recall with a learning-based approach? Current systems provide no answers to such questions.

Further, even when the user already knows what step to take, often there is also no guidance on how to do the step, e.g., current EM systems often provide a set of blockers/matchers, but do not tell the user how to select among them. As another example, there is currently no guidance on how to sample or label tuple pairs (Example 2).

**4. Use Human-in-the-Loop, Not Tools-in-the-Loop:**
Current DI systems often take a "human-in-the-loop" approach, where they try to *automate* a DI step *end-to-end*, allowing human feedback at only various execution points.

In practice, however, many DI steps are very messy, requiring multiple iterations involving many subjective judgments from human users. Very often, by working on them users gain a better understanding of the problem, then revise many decisions on the fly (see Example 2). As a result, many DI steps are still very difficult to automate, and are executed instead by *human users* in an ad-hoc fashion. For these steps, many users have indicated that what they need, first and foremost, is guidance on how to execute the steps end-to-end, then (semi-)automated tools to address the "pain points" during the execution.

EXAMPLE 3. *Continuing with Example 2, many users that we have met said they want a step-by-step guide on how to take a sample S, and then label S in a way that minimizes their effort.*

*Before labeling, they want to run a tool that processes S and highlights possible matching categories, so that they can develop the most comprehensive matching definition. For instance, users know that two companies with the same names and addresses should match. But the tool may show that S also contains many cases of companies with the same names but different addresses. This would force users to decide what to do with such cases. Upon closer inspection, users may find that these are branches of the same companies, and may decide that they should also match.*

*Then during the labeling process, if users must still revise the match definition, they want a tool that quickly flags already-labeled pairs in S that may need to be relabeled.*

Many current DI systems do not take the above "tools-in-the-loop" approach, where humans execute the "loop" and use tools to address the pain points, and thus often do not address the true pain points of real-world users.

**5. Blurring the Development and Production Stages:**
It is well-known that in practice users often execute a DI task in two stages: development and production, with very different challenges, e.g., maximizing accuracy vs. scaling, crash recovery, quality monitoring, etc. (see Example 1).

When using a current DI system, it is often not clear what support is there for each stage (and if the system distinguishes the stages at all). For example, many DI systems provide a way (e.g., a GUI) to specify a workflow then execute it on input data. It is not clear which stage this is intended for, most likely production (since it is too limited as a tool to find an accurate workflow for the development stage). But the development stage comes *first*. Users would need far more robust tools to help develop a good workflow, *before* they can even think about executing it in production. Such tools are missing from current DI systems.

**6. Not Designed from Scratch for Extendability:** In practice users often want to customize, extend, or patch a DI system. First, users often want to customize a generic DI system to a particular domain. Second, users may want to extend the system with latest technical advances, e.g., crowdsourcing, deep learning. Finally, users often have to write code, e.g., to implement a lacking functionality or combine system components. Writing "patching" code correctly in "one shot" (i.e., one iteration) is difficult. Hence, ideally such coding should be done in an interactive scripting environment, to enable rapid prototyping and iteration.

Few if any of the current DI systems are designed from scratch such that users can very easily customize, extend, and patch in many flexible ways. Most systems provide "hooks" at only certain points in the DI pipeline for adding limited new functionalities (e.g., a new blocker/matcher), and the vast majority of systems are not situated in an interactive scripting environment, making patching difficult.

**7. Not Designed for Collaborative Settings:** DI "in the wild" is surprisingly collaborative, e.g., multiple people (in different locations) trying to label, debug, and clean the data. Yet most current DI systems provide no or very limited capabilities for such collaboration.

## 3.2 Systems for Multiple DI Problems

Such a system jointly solves a set of DI tasks, e.g., data cleaning, schema matching and integration, then EM. This helps users solve the DI application seamlessly end-to-end (without having to switch among multiple systems), and enables runtime/accuracy optimization across tasks. Our experience suggests that these systems suffer from the following limitations.

(1) For each component DI task, these systems have the same problems as the systems for a single DI problems.

(2) As should be clear by now, building a system to solve a single DI task is already very complex. Trying to solve multiple such tasks (and accounting for the interactions among them) in the same system often exponentially magnifies the complexity.

(3) To manage this complexity, the solution for each component task is often "watered down", e.g., fewer tools are provided for both the development and production stages. This in turn makes the system less useful in practice.

(4) If users want to solve just 1-2 DI tasks, they still need to install and load the entire system, a cumbersome process.

(5) In many cases optimization across tasks (during production) does not work, because users want to execute the tasks one by one and materialize their outputs on disk for quality monitoring and crash recovery.

(6) Finally, such systems often handle only a pre-specified set of workflows that involves DI tasks from a pre-specified set. If users want to try a different workflow or need to handle an extra DI task, they need another system, and so end up combining multiple DI systems anyway.

## 4. THE PROPOSED AGENDA

To address the above limitations, we propose the following novel system building agenda:

- Build systems, each of which helps power users solve a single core DI problem, as software packages in the Python data ecosystem (or PyData for short).

- Foster PyDI, an ecosystem of such DI software packages as a part of PyData, focusing on how to combine such packages to jointly solve multiple DI problems.

- Extend PyDI to the collaborative/cloud/crowd/lay user settings.

We now motivate and discuss these directions.

## 4.1 Build Systems for Core DI Problems

In this direction we propose that we build DI systems that satisfy the following requirements (see Figure 2 for the proposed architecture).

### 4.1.1 Scope

**Each System Solves a Single Core DI Problem:** From Section 3, it is clear that, despite many years of R&D, we still have not been able to build "basic systems" that can effectively solve individual DI problems. Hence, we propose that we devote far more effort to building such systems. As we know much better how to build them, we can leverage that knowledge to build "composite systems" to jointly solve multiple DI problems (as will be discussed in Section 4.2).

Specifically, most DI applications involve a set of *core DI problems*: wrapper-based extraction, schema matching, entity matching, etc. We propose that we build "basic systems" each of which solves just one such core problem.

**Systems Target Power Users:** In the same vein, we still do not know how to build systems that effectively help *power users*: those that may not be DI experts but can code (e.g., data scientists). Hence, we propose that the "basic systems" should just target such power users for now. Later we can leverage these to build systems for *lay users* (e.g., those that cannot code), in a way analogous to building on assembly languages to develop higher-level CS languages.

### 4.1.2 Stages, Guides, and Tools

**Distinguish Development & Production Stages:** Following real-world practice, we propose that a user solve the core DI problem in two stages: developing an accurate DI workflow in the *development stage*, using data samples, then executing the workflow in the *production stage* on the entirety of data, focusing on scaling, crash recovery, quality monitoring, exception handling, etc.

**Develop How-To Guides for Both Stages:** For each stage we should develop a how-to guide, which should state as clearly as possible how the stage should be executed, step by step: which step should be first, which should be second, etc. For each step, the guide in turn should provide as detailed instructions as possible on how to execute it.

**Develop Tools for the Pain Points in the Guides:** How-to guides assume the human user will execute the steps. It is important to identify the pain points in this human-driven process, and develop (semi-)automated tools to reduce the human effort as much as possible.

**Guides & Tools Must Cover All Steps of DI Task:** To build truly practical systems, we must cover end to end. For example, for EM we cannot just do blocking and matching, as the vast majority of current EM works do. (This is akin to continually developing join algorithms without building the rest of the RDBMS.) We must cover all other steps, e.g., sampling, labeling, debugging, cleaning, as well.

### 4.1.3 Tools as Packages in the PyData Ecosystem

**Tools for Development Stage on Data Analysis Stack:** We observe that what users try to do in the development stage is very similar in nature to data analysis tasks. For example, creating EM rules can be viewed as analyzing data to discover accurate EM rules, and often requires users to perform tasks such as cleaning, visualizing, finding outliers.

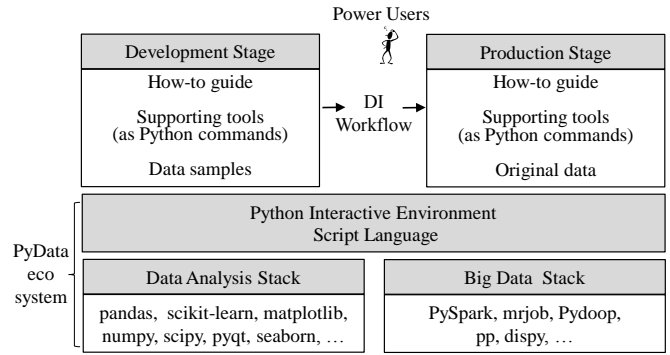

Figure 2: Proposed architecture for DI systems.

As a result, if we are to develop tools for this stage in isolation, within a stand-alone monolithic system, as current work has done, we would need to somehow provide a powerful data analysis environment, in order for these tools to be effective. This is clearly very difficult to do.

So instead, we propose that tools for this stage be developed on top of an open-source data analysis stack, so that they can take full advantage of all the data analysis tools already (or will be) available in that stack. In particular, two major data analysis stacks have recently been developed, based on R and Python. The Python stack for example includes the Python scripting language, numpy and scipy packages for numerical/array computing, pandas for relational data management, scikit-learn for machine learning, etc. More Python packages are being added all the time (e.g., PyPI, the largest Python package repository, contains 86,800+ packages as of August 2016). We propose to develop tools on this stack (then later build on it to develop tools for other stacks, e.g., R).

**Tools for Production Stage on Big Data Stack:** Similarly, we propose that tools for the production stage, where scaling is a major focus, be developed on top of the Python Big Data stack, which consists of packages to run MapReduce (e.g., Pydoop, mrjob), Spark (e.g., PySpark), and parallel/distributed computing in general (e.g., pp, dispy).

EXAMPLE 4. *We have built* Magellan, *a system that helps users solve the EM task described in Example 1. We provide detailed how-to guides, then develop tools for the pain points. These tools (a) take a sample from two tables A and B (ensuring a reasonable number of matches in the sample), (b) debug the blockers, (c) debug the labeling process, (d) select the best matcher, and (e) debug the matchers, among others. There are 104 Python commands that users can use. By leveraging 11 packages in the PyData ecosystem, we were able to build the current system (with a rich set of capabilities) quickly, with relatively little effort.*

EXAMPLE 5. *We have also built* Conform, *a DI system to normalize attribute values. Given a set S of values,* Conform *examines if any clustering algorithm that it has is likely to work well on S. If not,* Conform *helps user U manually normalize the values in S, using a detailed how-to guide and a GUI tool. Otherwise,* Conform *applies the clustering algorithm to S, then helps U manually "clean up" the output clusters, using a how-to guide. The system uses the pyqt package, among others, to quickly build GUI capabilities.*

### 4.1.4 Design Tools From Scratch for Interoperability

**Open-World versus Closed-World Systems:** As described, the proposed DI systems will be in the PyData ecosystem and expected to "play well" with other packages. We say that these systems are "open-world", in contrast to current stand-alone "closed-world" DI systems.

It is critical that we design these "open-world" DI systems from scratch for interoperability, so that they can easily exploit the full power of PyData and can be seamlessly combined to solve multiple DI problems. In particular, these systems should expect other systems (in the ecosystem) to be able to manipulate their own data, they may also be called upon by other systems to manipulate those systems' data, and they should be designed in a way that facilitates such interaction. This raises many interesting challenges, as we discuss in Section 5.

## 4.2 Foster a DI Ecosystem as a Part of PyData

By building open-world DI systems as described above, we are in effect building an *ecosystem* of interacting DI tools (as a part of the bigger PyData ecosystem). An important goal of our agenda is to grow this ecosystem, which we will call PyDI for short. To do so, we propose to study PyData and similar ecosystems, then apply the lessons to grow PyDI.

### 4.2.1 Study PyData and Similar Ecosystems

There is no denying that PyData has been very successful. We believe that our community should devote more effort to studying this and similar ecosystems (e.g., R). This can help us understand better the nature of certain data management problems (e.g., DI, data cleaning), promising solution approaches, and ways to foster similar ecosystems to solve those problems. Examples of issues we can explore include (but are not limited to) the following.

**What Do They Do?** The PyData community has been working on a wide variety of issues. First, they build tools to solve problems (e.g., Web crawling), implement cross-cutting techniques (e.g., learning), and help users manage their work (e.g., Jupyter notebook). Second, they develop extensive software infrastructure to build tools, and ways to manage/package/distribute tools. Third, they extensively educate developers/users, using books, tutorials, conferences, etc. Finally, they foster many players (companies, non-profits) to work on the above issues.

**Why Are They Successful?** Our experience suggests three main reasons. First, tools are often developed to address creators' pain points. Other users doing the same task often have the same pain points and thus find these tools useful. Second, creators consciously try to make tools easy to share, and much community effort has been spent on making popular tools easy to interoperate. Finally, tools are free and open-source, making it cheap and easy for a wide variety of users to use and adapt. (The extensive community effort to assist developers/users also helps a lot.)

**What Are Their Problems?** Despite their rapid growth, surprisingly there are still very few effective tools to do data wrangling (e.g., cleaning, IE, DI) in PyData, and very little guidance exists on how to solve these problems. In addition, building data intensive tools that interoperate raises many challenges, e.g., how to manage metadata/missing values/type mismatch across packages (see Section 5). Currently only some of these issues have been addressed, in an ad-hoc fashion.

### 4.2.2 Apply the Lessons Learned to Foster PyDI

It is clear that our community has much to contribute. In particular, growing PyDI will fill in the big data wrangling "gap" in PyData. But the PyData experience suggests that, to be successful, we must build tools that truly address user pain points, that are very easy to share and use, and that are open source. The system building methodology we have discussed so far (Section 4.1) tries to ensure these.

It is also clear that we cannot just build tools for DI problems (e.g., EM). Users also often need many tools to *manage their work*. The PyData community understands this and has developed many effective tools (e.g., Jupyter notebook/hub). We should develop similar tools for DI (e.g., loading/saving DI workflows, provenance, etc.). In addition, our community has recently developed many cross-cutting techniques (e.g., visualization, crowdsourcing), which would be highly desirable to add to PyDI.

Needless to say, we should also make sure that tools can be combined seamlessly to jointly solve multiple DI problems. Here the PyData experience suggests that it might not be sufficient to just rely on individual creators. Community players may have considerably more resources for cleaning/combining packages. So it is important that we foster such players (e.g., startups, non-profits, research labs, data science institutes, etc.).

Finally, just as in the PyData case, we should also work extensively on helping PyDI developers/users in terms of infrastructure, books, conferences, community resources, etc.

## 4.3 Build Collaborative/Cloud/Crowd/Lay User Versions of DI Systems

So far we have developed DI systems for a single power user, in his/her local environment. There are however increasingly many more DI settings, which can be characterized by *people* (e.g., power user, lay user, a team of users, etc.) and *technologies* (e.g., cloud, crowdsourcing, etc.). For instance, a team of scientists wants to solve a DI problem collaboratively, or a lay user wants to do cloud-based DI because he/she does not know how to run a local cluster.

To maximize the impacts of DI systems, we should also consider these settings. In particular, we briefly discuss three concrete settings below, and show that these settings can build on DI systems developed so far, but raise many additional R&D challenges.

**Systems for Lay Users:** A promising direction is to customize DI systems discussed so far for lay users, by adding GUIs/wizards/scripts as a layer on top. A lay-user action on a GUI, for example, is translated into commands in the underlying system (in a way analogous to translating a Java statement into assembly code). A key challenge is to build this top layer in a way that is easy for lay users to use yet maximizes the range of tasks that they can perform.

**Collaborative Systems:** Similarly, we can try to extend the DI systems discussed so far to collaborative settings. This raises many interesting challenges, e.g., how can users (who are often in different locations) collaboratively label a sample and converge to a matching definition along the way? How can they collaboratively debug, or clean the data? How can power users and lay users work together?

**Hands-Off DI:** These systems solve the entire DI task using crowdsourcing, requiring very little or no work from the user (i.e., the task's owner). For example, to solve the EM problem in Example 1, a user only needs to supply 2 positive and 2 negative examples, and an instruction to the crowd on how to match [7, 3]. Using such systems, an organization can solve far more DI tasks (by paying the crowd to solve them). Lay users can also easily solve DI tasks. For example, a user can upload two tables to a website, supply a few labeled examples, instructions to the crowd, and a credit card, the website simply enlists a crowd to match the two tables then returns the results to the user. As such, this is an example of a cloud/crowd/lay user DI system. Again, we can consider how to extend DI systems developed so far to build hands-off systems. A key challenge is to decide what crowd workers can do and how that can be translated into an accurate DI workflow in the underlying system.

## 5. ONGOING WORK AT WISCONSIN

We now describe ongoing work at Wisconsin based on the above agenda, and the main observations so far.

**The Proposed Agenda:** We found that this agenda could be used to effectively build a variety of DI systems. Back in 2014 we spent a year building an initial version of Magellan, a Java-based stand-alone EM system, following common practice: the system translates (GUI/command-based) user actions into a workflow of pre-defined operators, then optimizes and executes the workflow. We had serious difficulties trying to extend the system in a clean way to cope with the messiness of real-world EM tasks, where iterations/subjective decisions are the norm, and where exploratory actions (e.g., visualizing, debugging) are very common but it is not clear where to place them in the translate-optimize-execute-workflow paradigm.

Once we switched to the current agenda, these difficulties cleared up. We were able to proceed quickly, and to flexibly extend Magellan in many directions. Using PyData allowed us to quickly add a rich set of capabilities to the system. Besides EM, we found that the same methodology could also be used to effectively build systems for attribute value normalization (see Example 5) and string similarity joins. Finally, we are indeed able to extend Magellan to build a hands-off crowdsourced EM system (see more below).

**The DI Systems:** The DI systems we have built appear to be quite promising in helping users effectively solve DI tasks. In 2015 we asked 44 students in a data science class to apply Magellan to 24 diverse real-world EM tasks. They were able to follow the how-to guides to achieve high EM accuracy on all data sets (improving $F_1$ by as much as 72% compared to a baseline [10]). Various tools developed for Magellan (e.g., debuggers) proved highly effective in helping to reach this accuracy. The students also exploited a broad range of capabilities, e.g., cleaning, extraction, visualization, underscoring the importance of placing Magellan in an ecosystem that supplies these capabilities. Magellan has now been in production at three industrial partners (Johnson Control, WalmartLabs, and Marshfield Clinic) and successfully used to solve a range of EM problems. Our latest DI systems (string joins, value normalization) have also proven promising, but need to be evaluated more extensively. In addition, we have started developing BigGorilla (*biggorilla.org*, a joint project with Recruit Institute of Technology), a repository of DI tools for the PyData ecosystem. Finally, we have been building Corleone and Falcon, hands-off crowdsourced EM systems [7, 3], and have successfully used them to match tables of 1.8M-2.5M tuples at the cost of only $57-65. We are currently deploying them as a DI service on the cloud.

**The Challenges:** Building the above DI systems raises many challenges. Developing good how-to guides, even for the simple EM scenario of using supervised learning, turned out to be quite difficult (e.g., see Examples 2-3). Developing tools (e.g., to debug blockers/matchers and the labeling process), as well as developing collaborative/cloud/crowd/lay user systems, pose difficult research problems. Designing open-world systems (Section 4.1.4) also raises many issues [10]. For example, what kinds of data structures should a tool $T$ use to facilitate interoperability? How to manage metadata if any external tool can modify the data of $T$, potentially invalidating $T$'s metadata without $T$ knowing about it? How to manage missing values, data type mismatches, version incompatibilities, etc. across the packages?

This suggests that so far our community has only "skimmed" the surface of EM (and perhaps other DI tasks). Current EM work has focused mostly on the accuracy/cost of blockers/matchers. Expanding our focus to other parts of the EM pipeline, as suggested by this agenda, can raise many more interesting opportunities for R&D and practical impacts.

## 6. CONCLUSIONS

We argue that our community should devote far more effort to building DI systems, to truly advance the field. In this paper we have discussed the limitations of current DI systems, then proposed a novel system building agenda. Finally, we have described ongoing work at Wisconsin, which shows the promise of this agenda. More details about this initial work can be found at sites.google.com/site/anhaidgroup.